\documentclass[a4paper,11pt]{article}
\usepackage[utf8]{inputenc}

\usepackage{amsfonts}
\usepackage{amssymb}
\usepackage{graphicx}
\usepackage{amsmath}
\usepackage{enumerate}
\usepackage{color}
 \usepackage{multirow}
\usepackage{float}

\RequirePackage{mathrsfs} \RequirePackage[sc]{mathpazo}
\RequirePackage{wasysym} \RequirePackage{setspace}\textheight=650pt
\title{ {\bf On  Heat  Properties of   AdS Black Holes in Higher Dimensions}}
\author{A. Belhaj$^{1,2}$, M. Chabab$^2$, H. EL Moumni$^2$, K. Masmar$^2$, M.  B. Sedra$^{3,4}$, A. Segui$^{5}$ \\
\\
{\small $^{1}$D\'epartement de Physique, Facult\'e
Polydisciplinaire, Universit\'e Sultan
 Moulay Slimane, B\'eni Mellal,  Morocco. } \\
{\small $^{2}$High Energy Physics and Astrophysics Laboratory, FSSM,
 \small Cadi Ayyad University, Marrakesh, Morocco.
} \\
{\small $^{3}$  D\'{e}partement de Physique, LHESIR, Facult\'{e} des
Sciences, Universit\'{e} Ibn Tofail,
 K\'{e}nitra, Morocco.} \\
 {\small \hspace{-1.5 cm}$^{4}$ Universit\'e Mohammed Premier,
  Ecole Nationale des Sciences Appliqu\'ees, BP : 3, Ajdir, 32003, Al Hoceima, Morocco.}\\
  {\small \hspace{-1.5 cm}$^{5}$  Departamento de F\'isica Te\'orica,
Universidad de Zaragoza, E-50009-Zaragoza, Spain}
 }
\date{}

\begin{document} \maketitle
\begin{abstract}
 We investigate the heat properties of AdS Black Holes in higher dimensions.
  We consider the study  of  the corresponding thermodynamical properties
  including the heat capacity explored in the determination of    the  black
   hole stability. In particular,  we  compute the heat latent.  To overcome the instability problem,
 the Maxwell construction,  in the $(T,S)$-plane, is elaborated. This
 method is  used to modify the the Hawking-Page phase structure by
 removing  the negative heat capacity regions. 
 Then, we discuss the thermodynamic cycle and
 the heat engines using the way based on the extraction of the work from a black hole solution.
\end{abstract}
\newpage
\section{Introduction}
Recently, many efforts have been devoted to study the
thermodynamical properties of  black holes, in connection with
higher dimensional supergravity models \cite{01,02}. These
properties  have been extensively studied via different methods
including numerical computation  using various codes \cite{04}. In
fact, several  models based on  mathematical methods have been
explored to study critical behaviors  of  black holes having
different geometrical configurations in arbitrary dimensions. 
A particular emphasis has been put on  AdS black
holes \cite{Kastor:2009wy,30,a1,4,5,50,6}. More precisely, a
nice interplay between the behaviors of the RN-AdS black hole
systems and the Van der Waals fluids has been shown \cite{KM,chin1,our,our1,our2}.
In this context, several landmarks of statistical liquid-gas systems, such as the P-V criticality, the Gibbs free energy, the first order
phase transition and the behavior near the critical points  have been derived.
 Also, in arbitrary dimensions of the spacetime, the authors \cite{KM,
our} studied the critical behaviors of charged RN-AdS black holes.  Extension to other solutions 
considered as a subject of interest in gravity theory, has  also been performed  and their corresponding phase
transitions and statistical  properties  have been investigated
using different approaches  \cite{Spallucci:2013osa,Spallucci:2013jja,hasan,Zhao:2014eja}. More recently, some authors have worked out the heat properties of AdS charged black holes and their solutions in four dimensions \cite{Dolan,Cliff}.

The aim in this paper is to 
reconsider the heat properties of AdS black holes in higher
spacetime dimensions. More precisely,  we will study the
corresponding thermodynamical properties including  the sign of the
heat capacity explored when discussing  the stability problem.
In particular,  we will derive the expression of the latent heat 
considered as a trivial consequence of the Hawking-Page phase
transition. To overcome the instability  problem, the Maxwell
construction  in $(T,S)$ plane is then elaborated to modify 
the Hawking-Page phase structure \cite{Zhao:2014eja,Dolan}. Finally, we will discuss the thermodynamic
cycle and the holographic heat engines.

The paper is organized as follows. In section $2$, we reconsider the study of thermodynamics of 
AdS black holes along with the latent heat. Section $3$ is devoted to Maxwell's construction of
higher dimensional AdS black holes. In section $4$, we discuss the thermodynamical cycle and
holographic heat engines. Finally, section $5$ contains our conclusions.

\section{Thermodynamics and latent heat }
This section concerns  the study of the latent heat properties of Ads black holes in higher dimensions.  This investigation  could be
supported by the existence of higher dimensional supergravity
theories including  superstring models, M-theory,  and related
topics. Here,  we consider a non-rotating, neutral,
asymptotically anti-de Sitter black holes in high dimensions spacetime $n\geq 4$.   The
corresponding metric solution  reads as
\begin{equation}
ds^{2}=-f(r)dt^{2}+\frac{dr^{2}}{f(r)}+r^{2}d\Omega_{n-2}^{2}
\label{metric}
\end{equation}
where $d\Omega _{n-2}^{2}$  represents  the volume of
$(n-2)$-dimensional sphere.  This solution  is characterized
 by the function $f(r)$ taking  the following general form
\begin{equation} \label{2}
f(r)=1-\frac{2M}{r^{(n-3)}}+\frac{r^{2}}{\ell ^{2}}.
\end{equation}
It is worth  recalling  that the parameter $M$  indicates  the ADM
mass of such a  black hole solution while its  horizon radius $r_+$  can
be identified to the largest real root of $f (r) = 0$. These two
parameters are linked via the relation
\begin{equation}
M=\frac{ r_+^{n-3}}{2} \left(\frac{r_+^2}{\ell^2}+1\right).
\end{equation}
In the non-rotating AdS black holes, $M$ should be  interpreted as
an enthalpy \cite{Kastor:2009wy} which can be written as follows
\begin{equation}\label{1}
 H(S,P)=\frac{1}{2} \left(\frac{4S}{\omega
   }\right)^{\frac{n-3}{n-2}} \left(\frac{16\pi P
   }{(n-2) (n-1)}\left(\frac{4S}{\omega
   }\right)^{\frac{2}{n-2}}+1\right).
\end{equation}
where  the Bekenstein-Hawking entropy is  given in terms of  the
horizon
\begin{equation}
 S=\frac{\omega}{4} r_+^{n-2}.
\end{equation}
In this equation,  the quantity $\omega$  reads as
\begin{equation}
 \omega=\frac{2 \pi ^{\frac{n-1}{2}}}{\Gamma
 \left(\frac{n-1}{2}\right)}.
\end{equation}
In fact,  many other thermodynamical  quantities  can be also computed
using similar technics. Indeed,  the pressure  can be  associated
with the cosmological constant $\Lambda$ and they are related as  $P
=-\frac{\Lambda}{8\pi}=\frac{(n-2)(n-1)}{16 \ell^2 \pi}$, whereas 
the temperature is  given  in terms of the horizon
radius $r_+$ via the following form
\begin{equation}\label{T}
 T=\frac{f'(r_+)}{4\pi}=-\frac{n(n-5)-2 r_+^2 \Lambda +6}{4 \pi
 r_+(2-n)}.
\end{equation}
By combining thermodynamical relations, the temperature can be re-written as 
\begin{equation}\label{TS}
 T=\left(\frac{\partial H}{\partial S} \right)_P=\frac{4^{\frac{1}{2-n}} \left(n^2-5 n+6\right) \left(\frac{S}{\omega }\right)^{\frac{1}{2-n}}+\pi  2^{\frac{2}{n-2}+4} P
   \left(\frac{S}{\omega }\right)^{\frac{1}{n-2}}}{4 \pi  (n-2)}.\end{equation}
One can see that the temperature presents a  minimum associated
with  the following entropy value
   \begin{equation}\label{smin}
    S_{T_{min}}=\pi ^{\frac{2-n}{2}} \omega  \left(\frac{2^{\frac{2 (1-n)}{n-2}} \sqrt{n^2-5
   n+6}}{\sqrt{P}}\right)^{n-2}.
   \end{equation}
 This minimum is given by
\begin{equation}\label{tmin}
 T_{min}=\frac{2 (n-3) \sqrt{P}}{\sqrt{\pi } \sqrt{(n-3) (n-2)}}.
\end{equation}

A similar computation shows that  the heat capacity can be expressed as

\begin{equation}\label{eq11}
C_p=\left(T\frac{\partial S}{\partial T}\right)_P=\frac{(n-2) S \left(\pi \; 2^{\frac{4}{n-2}+4} P \left(\frac{S}{\omega }\right)^{\frac{2}{n-2}}+n^2-5 n+6\right)}{\pi\;  2^{\frac{4}{n-2}+4} P
   \left(\frac{S}{\omega }\right)^{\frac{2}{n-2}}-n^2+5 n-6}
\end{equation}

Note that this quantity  is negative  for $S<S_{T_{min}}$. It becomes
positive for $S>S_{T_{min}}$, but diverges  at  $S=S_{T_{min}}$.

 Besides, recalling that the variation of Gibbs free energy  $G$ is
 \begin{equation}\label{vargibbs}
 dG=-SdT+PdV,
 \end{equation}

and  knowing that $G$ is the Legendre transform
of the enthalpy, one finds 
\begin{equation}\label{gibbs}
 G=H-TS=\frac{4^{\frac{n-1}{2-n}}}{\pi } S(T,P) \left(\frac{S(T,P)}{\omega }\right)^{\frac{1}{2-n}} \left(1-\frac{\pi
   4^{\frac{2 (1-n)}{2-n}} P \left(\frac{S(T,P)}{\omega }\right)^{\frac{2}{n-2}}}{n^2-3
   n+2}\right).
\end{equation}
where the entropy function given in  terms of $T$ and $P$
reads as

\begin{eqnarray}\label{eq13}\nonumber
S(T,P)&=&2^{4-5 n}    (\pi P)^{2-n} \omega\left[   2^{\frac{2 n}{n-2}} (n-2)
  \pi T \right.\\ &\pm&\left.\sqrt{\pi } \sqrt{(n-2) \left(\pi  2^{\frac{4
   n}{n-2}} (n-2) T^2-2^{\frac{8}{n-2}+6} (n-3) P\right)}\right]^{n-2}
\end{eqnarray}


Let 
\begin{equation}
\mathcal{B}=\frac{(n^2-3
   n+2)  \omega ^{\frac{2}{n-2}}}{\pi \;
   4^{\frac{2 (1-n)}{2-n}}}
\end{equation}

Then, from Eq.(\ref{gibbs}) we see that  for  $PS^{\frac{2}{n-2}}>\mathcal{B}$ the Gibbs free energy is negative,  hence the black hole is a more stable thermodynamical  configuration than the Anti-de Sitter one.

For $PS^{\frac{2}{n-2}}<\mathcal{B}$, however, the  pure AdS space-time is the more
stable which means that the  black hole with $S<\left(\frac{\mathcal{B}}{P}\right)^\frac{n-2}{2}$
will  evaporate. This is associated with the line where the Hawking-Page
transition occurs. This line, dubbed coexistence line, is defined by the following equation
\begin{equation}\label{co1}
 S=4^{1-n} \left(n^2-3 n+2\right)^{\frac{n}{2}-1} \pi ^{1-\frac{n}{2}} \omega
 P^{1-\frac{n}{2}}.
\end{equation}
 where the pressure $P|_{coexistance}$ can be computed  in terms of the temperature for any
 dimension thanks to equations   (\ref{eq13})  and (\ref{co1}), 
\begin{equation}\label{co2}
P|_{coexistance}=\left\{\begin{array}{cc}
\frac{3 \pi  T^2}{8},& n=4 \\
 \frac{2 \left(32-85 \sqrt[3]{2}+168\ 2^{2/3}\right) \pi  T^2}{1849},&n=5 \\
 \frac{80 \pi  T^2}{529},&n=6  \\
 \vdots\\
 \frac{567 \pi T^2}{6241},&n=10.\\
 \vdots
\end{array}\right.
\end{equation}

\begin{center}
\begin{figure}[!ht]
\begin{tabbing}
\hspace{9cm}\=\kill
\includegraphics[scale=.7]{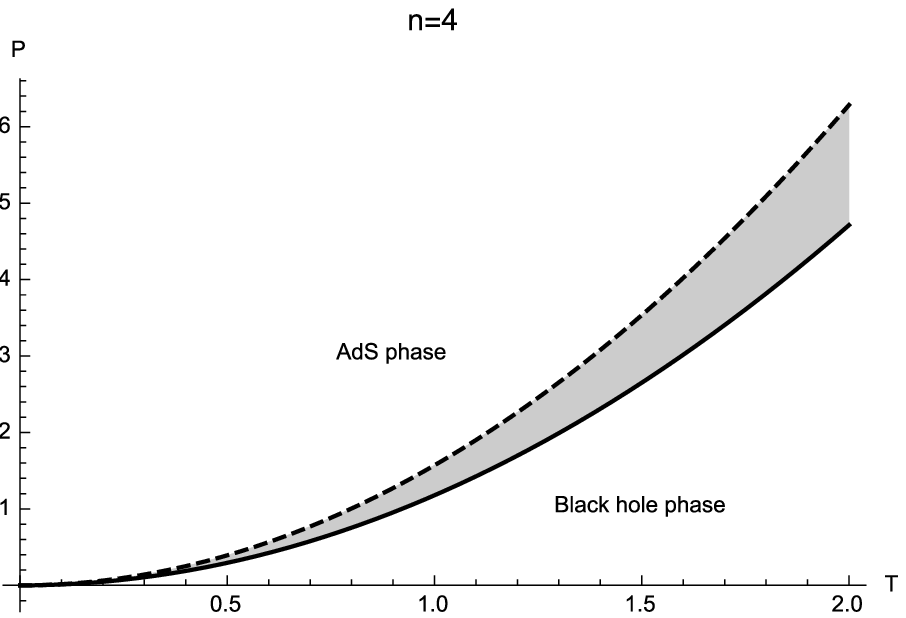}\>\includegraphics[scale=.7]{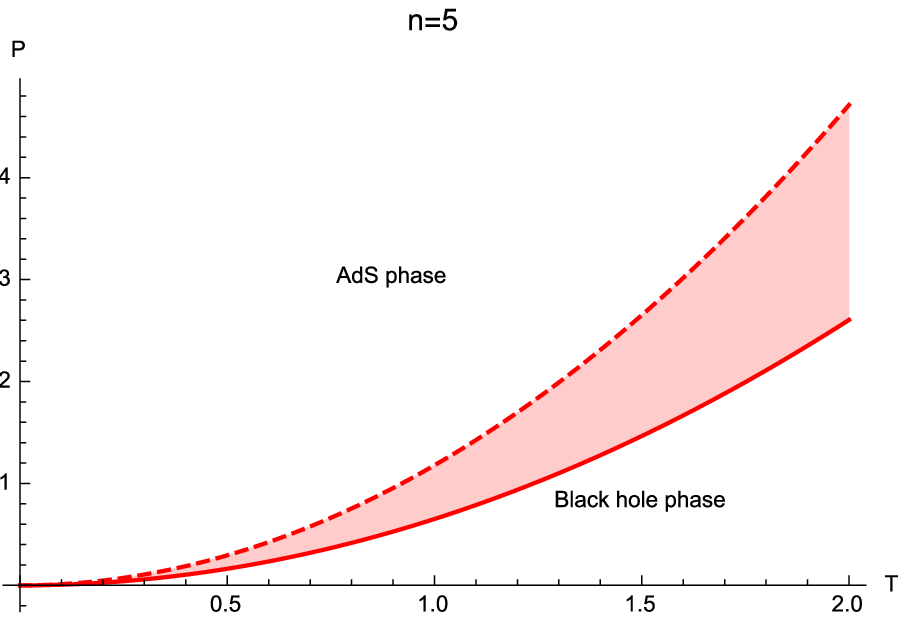} \\
\includegraphics[scale=0.7]{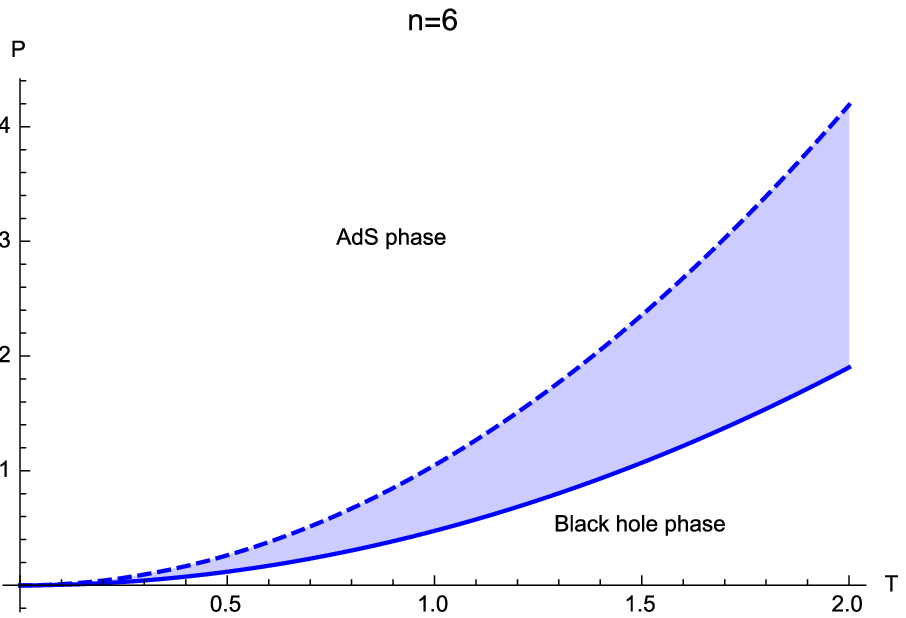}\> \includegraphics[scale=0.7]{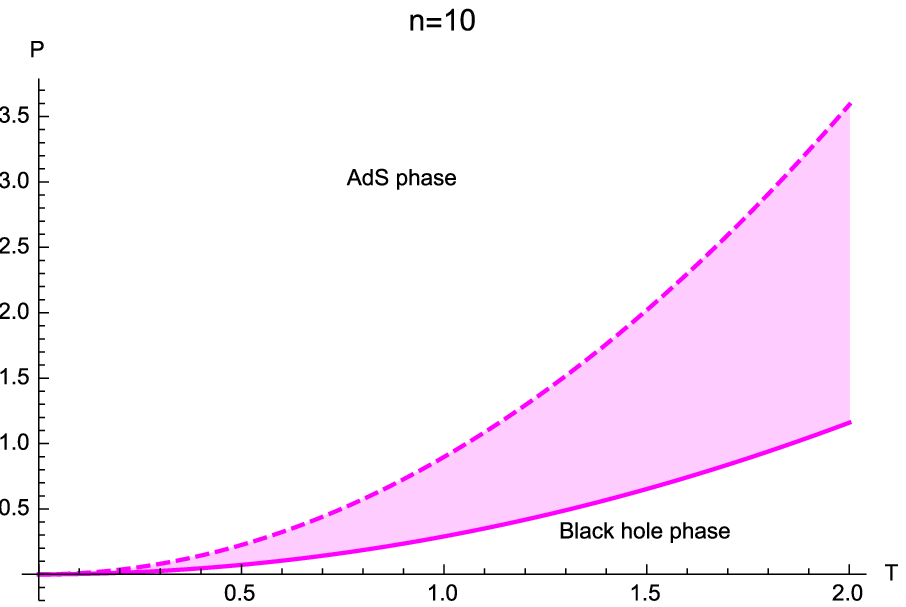}\\
\end{tabbing}
\vspace*{-.2cm} \caption{Phase diagram for higher
dimensional AdS black holes. The coexistence line of the
AdS-Radiation / Black hole phase transition of  such a  system in
$(P,T)$ plane.} \label{fig1}
\end{figure}
\end{center}

This is illustrated in Fig.$\ref{fig1}$ which clearly indicates the existence of the two phases.

This figure  shows the coexistence curve of the Hawking-Page phase
transition, represented by  the lower (solid) line. It also indicate
that  the heat capacity diverges on the upper (dashed) line while the
lower branch of the free energy goes  to minus infinity on the line
given by $P = 0$.

\begin{center}
\begin{figure}[!ht]
\begin{tabbing}
\hspace{8.0cm}\=\kill
\includegraphics[scale=.8]{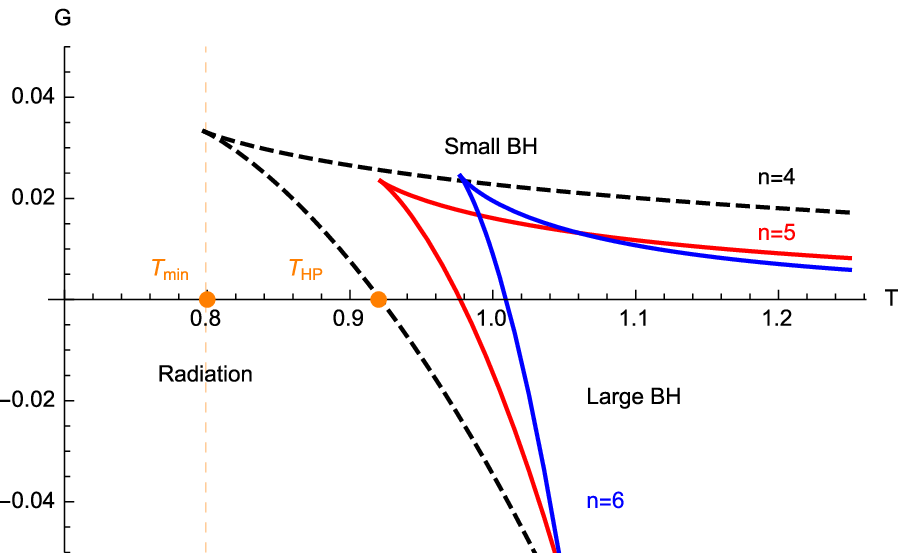}\>\includegraphics[scale=.8]{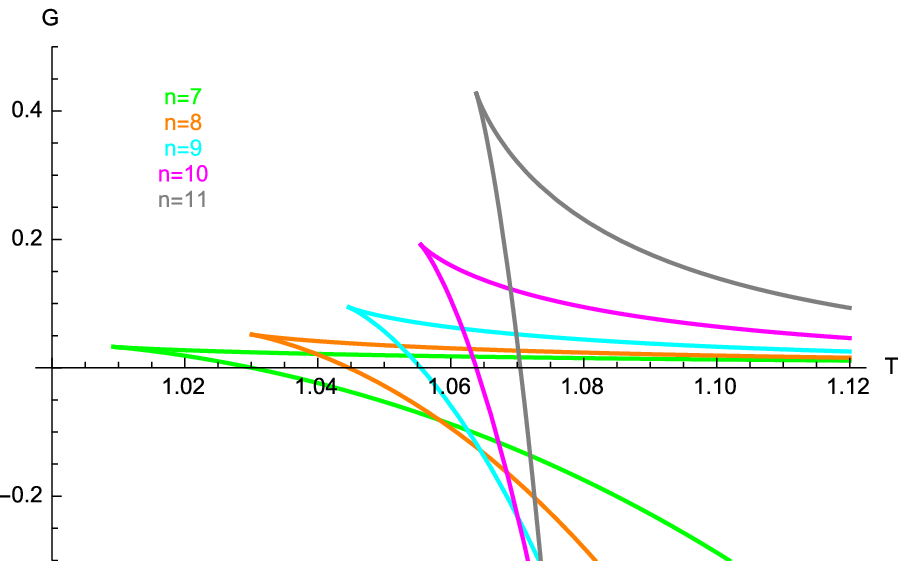} \\
\end{tabbing}
\vspace*{-.2cm} \caption{Left: The Gibbs free energy  as a function
of temperature at fixed $P=1$  for  $4\leq n\leq 6$.
 Right:  Higher dimensional  cases  associated with    $7\leq n\leq 11$ Schwarzschild-AdS black hole.}
\label{fig2}
\end{figure}
\end{center}

Next, we would like to address the stability issue of such solutions. To do so,  we plot in  Fig.$\ref{fig2}$ the
behavior of the Gibbs free energy $G$ in terms of the temperature
$T$ at fixed pressure.

 One can notice a minimum temperature $T_{min}$  for
which no black holes with $T < T_{min}$ can survive.  However,  above  this
temperature, two branches of  the black holes appear.  In
fact, the upper branch describes an  unstable small
(Schwarzschild-like) black hole involving a
 negative specific heat.  For ($T>T_{min}$),   the black holes, at lower branch,
 become stable with  positive specific heat.
 In addition since the  Hawking-Page temperature $T_{HP}$ is associated with vanishing values of
 the Gibbs free energy, then the black hole Gibbs free energy becomes negative
for  $T>T_{HP}$. In fact,  at $T = T_{HP}$, a first order
Hawking-Page  phase transition shows up between the  thermal radiations and large black holes as reported in \cite{14Dolan,15Dolan,Dolan}.

Moreover,  from Fig.\ref{fig1} we also note a jump in
entropy which becomes more relevant in terms of  the dimension of the space time
as shown in the following equation
\begin{equation}
 \Delta S=4^{1-n} \left(n^2-3 n+2\right)^{\frac{n}{2}-1} \pi ^{1-\frac{n}{2}} \omega
 P^{1-\frac{n}{2}}.
\end{equation}
Notice that  the latent heat of the black hole which nucleates from anti-de Sitter space time can be computed using
the following  thermodynamical  expression
\begin{equation}
 L=T\Delta S.
\end{equation}
Indeed, using Eq. (2.8),  the calculation results in
\begin{equation}
 L=\frac{2^{-\frac{n}{n-2}} (n-2) \omega }{\pi } \left(4^{1-n} \pi ^{1-\frac{n}{2}}
   \left(\frac{n^2-3 n+2}{P}\right)^{\frac{n-2}{2}}\right)^{\frac{n-3}{n-2}}
\end{equation}
  which is equal to the mass on the coexistence curve in the black hole phase.  Here, note that for  $n=4$,
we reproduce  the result reported in  \cite{Dolan}. It turns out that the latent heat is nonzero for any finite
 $T$ and  vanishes  for very large values of $T$. In fact,  in the case of  asymptotically  flat space-time with $P=0$,
  the latent heat becomes infinite which means that the black hole cannot nucleate in Minkowski space \cite{Dolan}.

 It is clear now that the sign  of the heat capacity plays an important role in the determination of
 the stability of the black hole.  More precisely,   its negative values render the black hole
 unstable. In the next section we will show how to overcome this problem, by using the Maxwell
construction to modify the Hawking-Page phase structure.

\section{On Maxwell's construction of high dimensional  AdS black holes}
\begin{figure}[!ht]
\begin{center}
\includegraphics[scale=1]{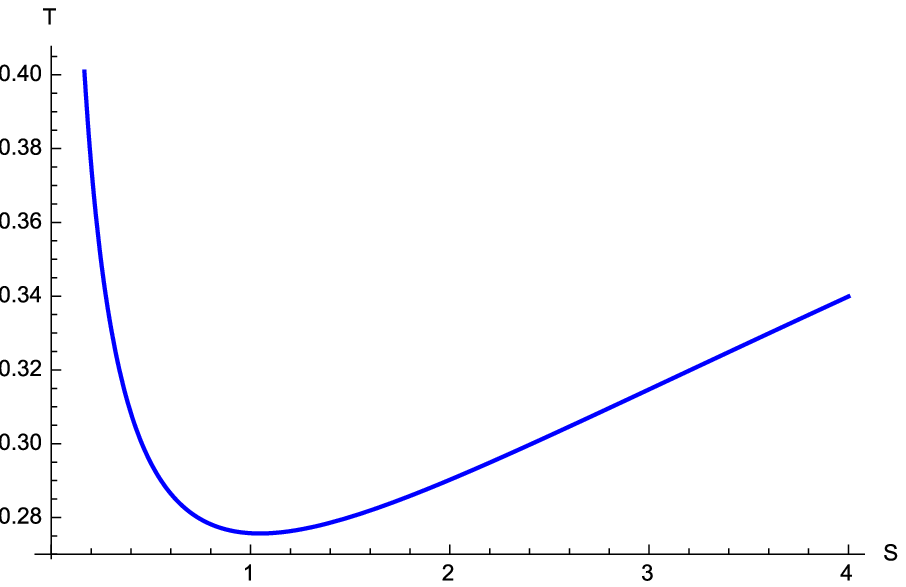}
\end{center}
 \caption{The temperature as function of the entropy in four dimension, with $P=1$.} \label{fig3} \vspace*{-.2cm}
\end{figure}

\begin{center}
\begin{figure}[!Ht]
\begin{tabbing}
\hspace{10cm}\=\kill
\includegraphics[scale=.66]{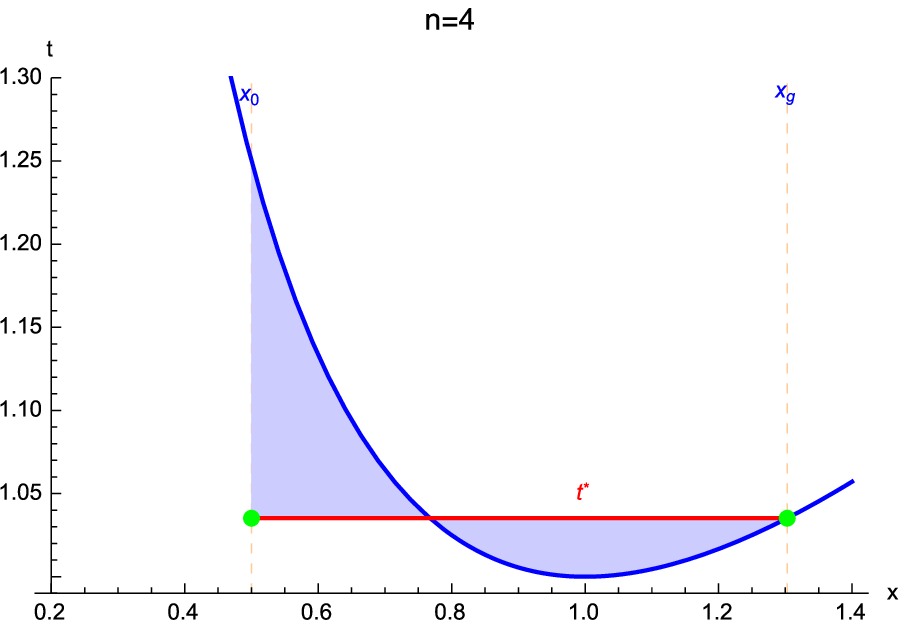}\>\includegraphics[scale=.66]{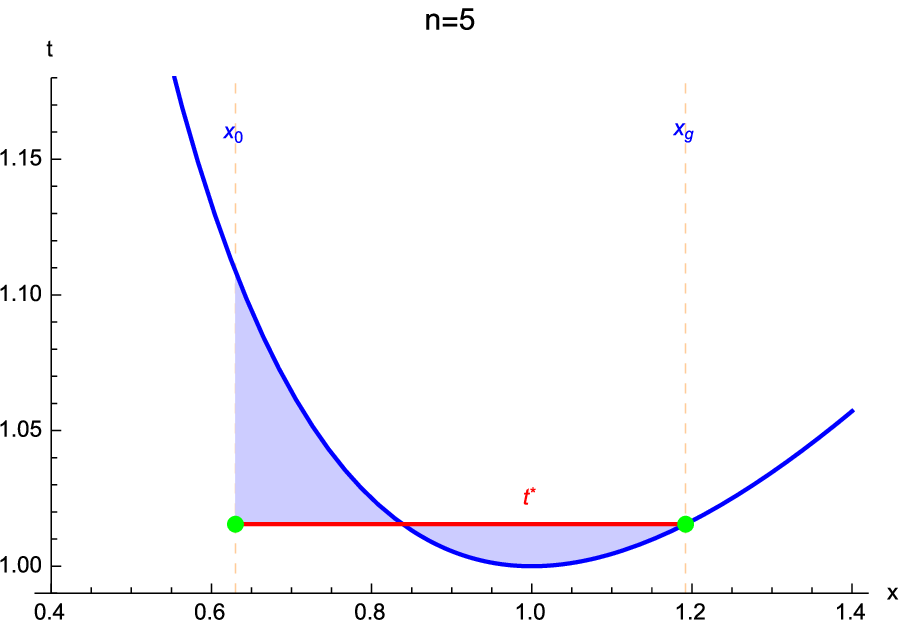} \\
\includegraphics[scale=0.66]{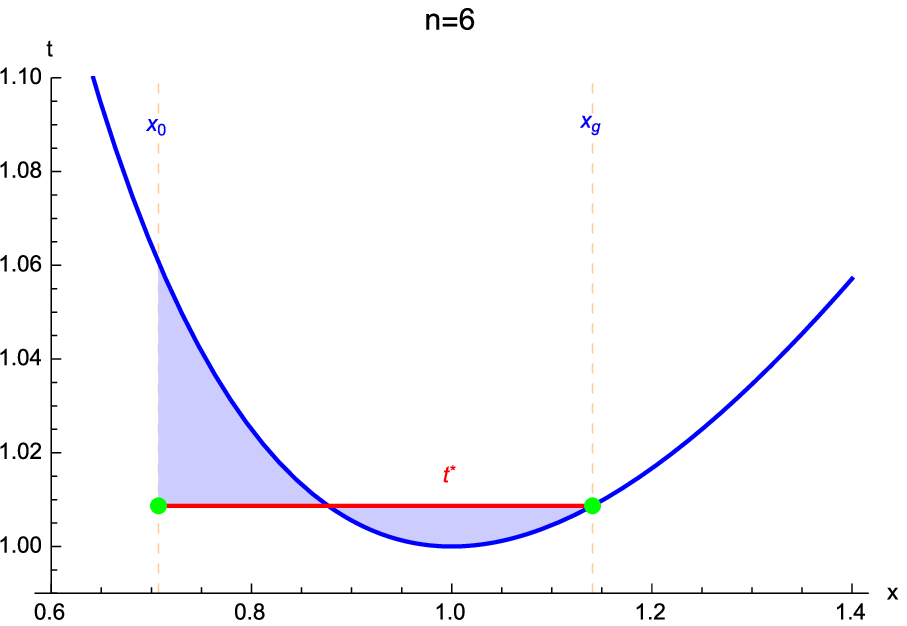}\> \includegraphics[scale=0.66]{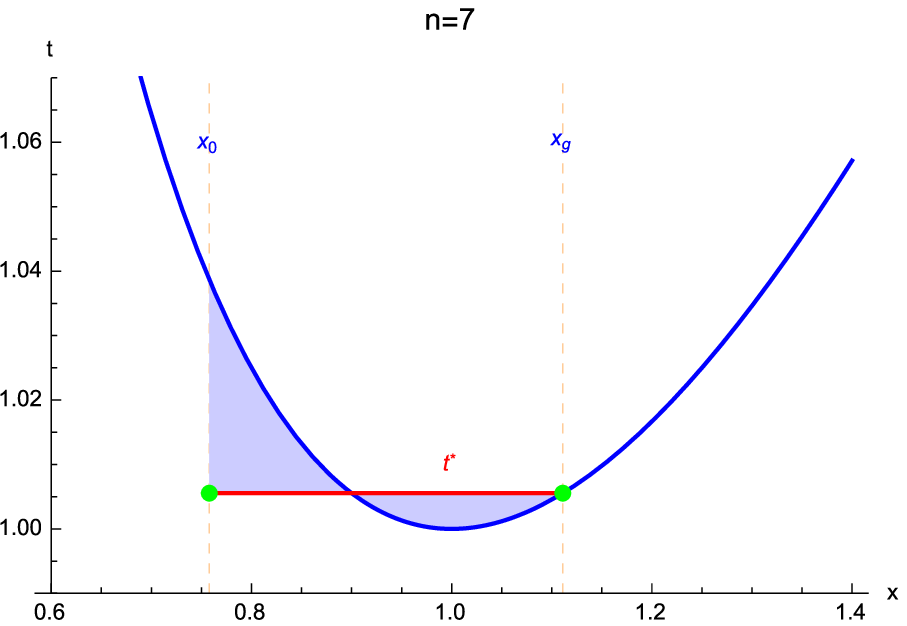}\\
\includegraphics[scale=0.66]{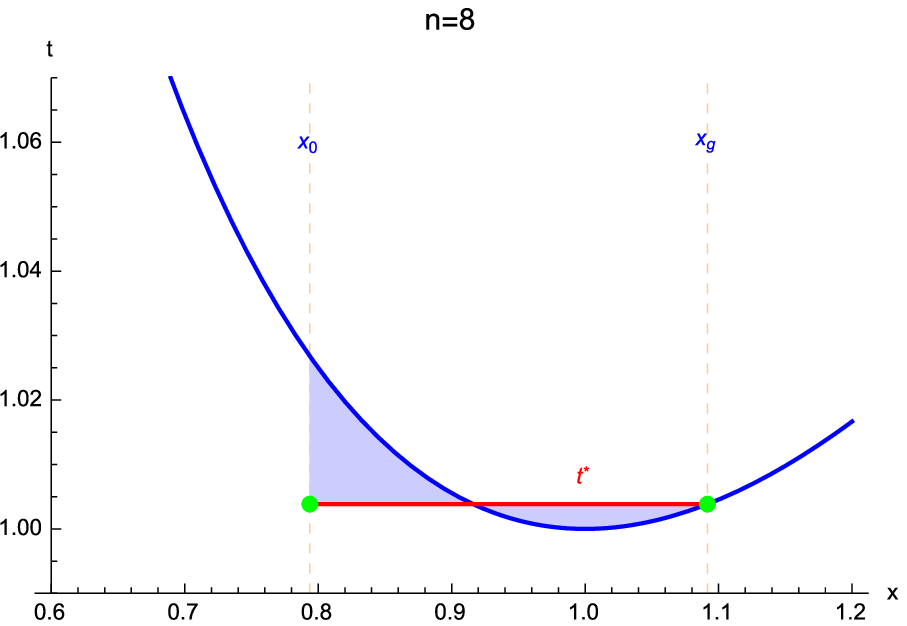}\>\includegraphics[scale=0.66]{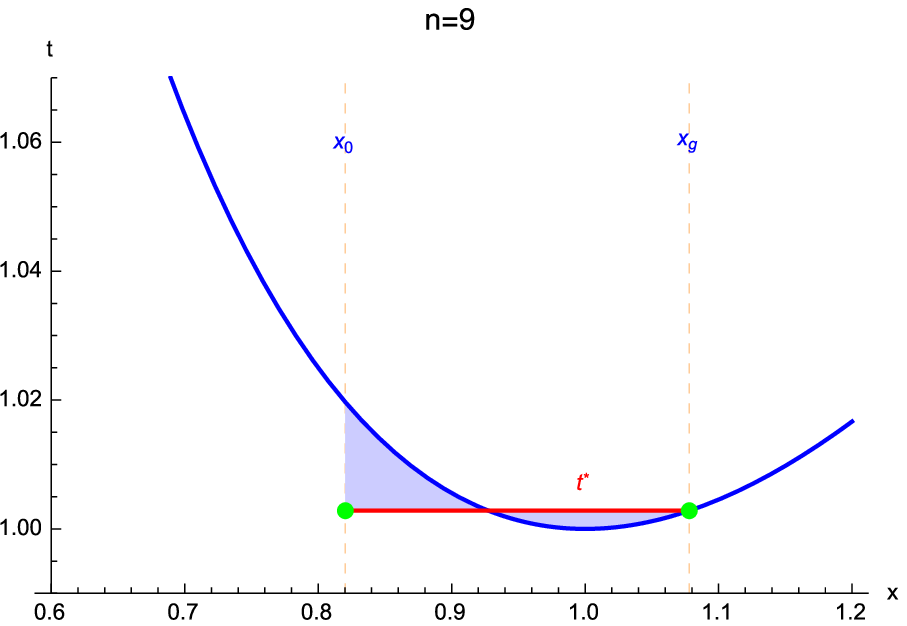} \\
\includegraphics[scale=0.66]{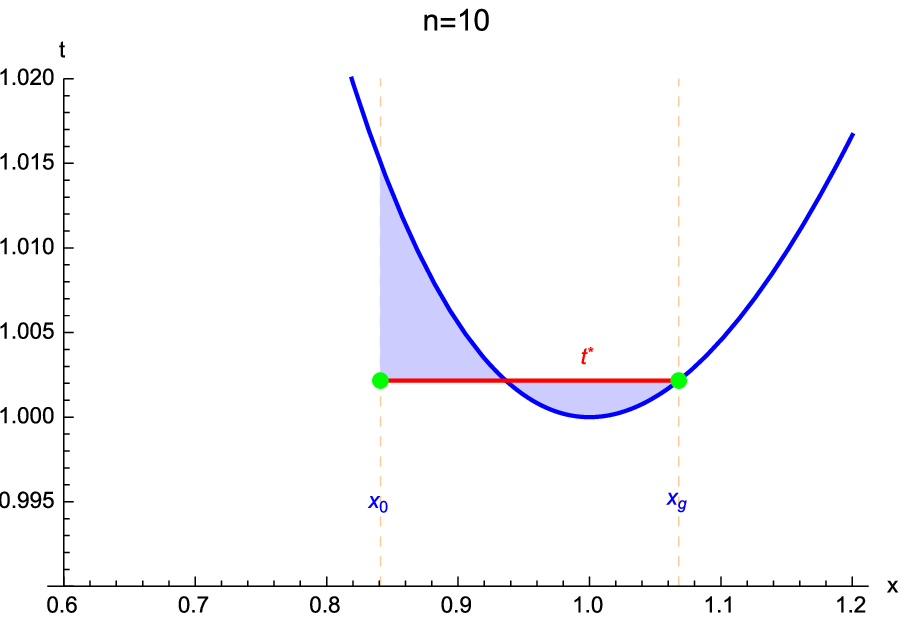}\>\includegraphics[scale=0.66]{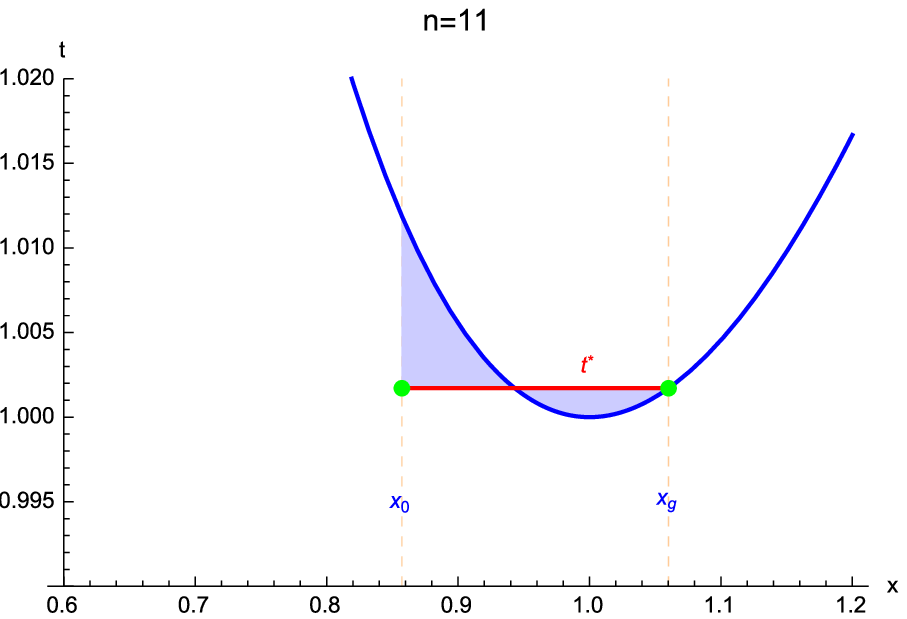} \\
\end{tabbing}
\caption{The $t-x$ diagrams  for space dimension $n$ between $4$ and $11$.
$t^\star$ and $x_{0,g}$ are given in table (1).}
\label{fig4}
\end{figure}
\end{center}

In this  section, we investigate the corresponding Maxwell's
construction. To do so, we  first recall that the equal area law was
introduced by Maxwell in order to explain the experimental behaviors
of real fluids. Usually, this construction is elaborated in the
$(P,V)$ plane while keeping constant temperature
  \cite{Spallucci:2013osa,hasan,Zhao:2014eja}.
 However, fixing   the pressure  in (\ref{vargibbs}), 
   such a  construction can also be done in
the $(T,S)$ plane. For Schwarzschild-AdS black hole, the choice of
this plane  has been explained in many papers including
\cite{14Dolan,Spallucci:2013osa,Spallucci:2013jja}. The starting
point is the temperature as a un function of the entropy given by
Eq.(\ref{TS}).  Then, we  plot    this function   in
Fig.$\ref{fig3}$.
It is observed that  this  function involves  minimums   at
$S_{T_{min}}$ and $T_{min}$ given by Eq.(\ref{smin})  and
Eq.(\ref{tmin}) respectively.   For any value of the cosmological
constant, these quantities produce  the following  reduced forms
\begin{eqnarray}
t&=& \frac{1}{2} s^{\frac{1}{n-2}}+\frac{1}{2}\;\frac{1}{s^{\frac{1}{n-2}}},\\
s&\equiv& \frac{S}{S_{T_{min}}} \quad and  \quad  t \equiv
\frac{T}{T_{min}}.
\end{eqnarray}
It is recalled that  the Maxwell area law  can be obtained  using
the fact the Gibbs free energy is the same for coexisting  black
holes.  Exploring  Gibbs free energy (\ref{gibbs})
\begin{equation}
\Delta G_{0,g}=- \int_1 ^2 SdT=0,
\end{equation}
\begin{equation}
T^\star \Delta S_{0,g}=\int_0 ^g TdS,
\end{equation}
where $T^\star$ is the temperature of the equal area isotherm. The
equal area law, in the reduced variables,   gives
 the entropy of the liquid and gaseous phases.  This solves
  the following equations
  \begin{equation}
\left\{\begin{array}{c}t=\frac{1}{2 s^{\frac{1}{n-2}}
 }\left(s^{\frac{2}{n-2}}+1\right) \\t^\star=\frac{(n-2)^2}{(n-3) (n-1)}
  \frac{s^{\frac{n-3}{n-2}}-s_0^{\frac{n-3}{n-2}}}{s_0-s}.
\end{array}\right.
\end{equation}
Introducing  new variable $x\equiv s^{\frac{1}{n-2}}$,  we get
the following equation
\begin{eqnarray}\label{poly}\nonumber
& &2^{\frac{2}{n-2}+2} (n-3) x^{n+2}-2^{\frac{2}{n-2}+2} (n-1) x^n-2^{\frac{2}{n-2}} (n-3)
   (n-1) x^4\\ &+&\left(n \left(2^{\frac{4}{n-2}}
   (n-3)+n-5\right)+2^{\frac{n+2}{n-2}}+6\right) x^3
   -2^{\frac{2}{n-2}} (n-3) (n-1) x^2=0.
\end{eqnarray}
 The solutions of this equation  associated
with each dimension are listed in Table $1$,

\begin{table}[!ht]
\begin{center}\begin{tabular}{|c|c|c|c|c|c|}\hline n & $x_0$ & $x_g$ & $s_0$ &$ s_g$ & $t^\star$  \\\hline
 4 & 0.50000& 1.30277 & 0.25 & 1.69722 & 1.03518 \\\hline
 5 & 0.62996 & 1.19213 & 0.25 & 1.69424 & 1.01548 \\\hline
 6 & 0.70710 & 1.14071 & 0.25 & 1.69319 & 1.00868\\\hline
  7 & 0.75785 & 1.11100 & 0.25 & 1.69271 & 1.00555\\\hline
  8 & 0.79370 & 1.09165 & 0.25 & 1.69244 & 1.00385 \\\hline
  9 & 0.82033 & 1.07805 & 0.25 & 1.69229 & 1.00283\\\hline
   10 & 0.84089 & 1.06796 & 0.25 & 1.69218& 1.00216 \\\hline
    11 & 0.85724& 1.06018 & 0.25 & 1.69211& 1.00171 \\\hline
    \end{tabular} \caption{roots of the polynomial form and the corresponding entropy and temperature. }
\end{center}
\label{tab}
\end{table}
We should  eliminate  the  states  corresponding to  either
 complex or negatives values since they have  no   physical meaning. In the $(T,S)$ plane (or equivalently in $(t,x)$),
 the system involves similar behaviors  associated with  the unstable (unphysical) part of the Van der Waals' picture in
   the $(P,V)$ plane. Roughly, in $\ref{fig4}$ we show the Maxwell's equal area in the $(t,x)$-plane for high
dimensional  Schwarzschild-Ads black hole.

It is observed  that  a pure radiation phase can  survive beyond
$T_{min}$ up to the higher temperature  given by
 the isotherm $T^\star=t^\star T_{min}$. For  $T=T^\star$,
the  black holes with  different entropy values  have the same
free energy. They are  more stable than the pure radiation. For
$T>T_{min}$, there exists a single and stable black hole with a
positive heat capacity.   When we go back to four dimensions,  we recover  the
results reported in \cite{Spallucci:2013osa}  and \cite{Spallucci:2013jja}
 describing neutral case.

\section{Thermodynamic cycle and the heat engines}
Having discussed  some  thermodynamical properties of the
Schwarzschild-AdS black hole including the thermodynamical quantities
associated with  stability and  phases transitions,  we pave the way to the study of
 the corresponding work from  the heat energy according to  a cycle between two  sources (cold/hot)
with temperature $T_C$ and $T_H$  respectively. Then, we make
contact with  the Carnot cycle defined as a  simple cycle described by two isobars and two
   isochores as in \cite{Cliff}. This is illustrated in Fig. 5.
 \begin{center}
\begin{figure}[!ht]
\begin{center}
\includegraphics[scale=1]{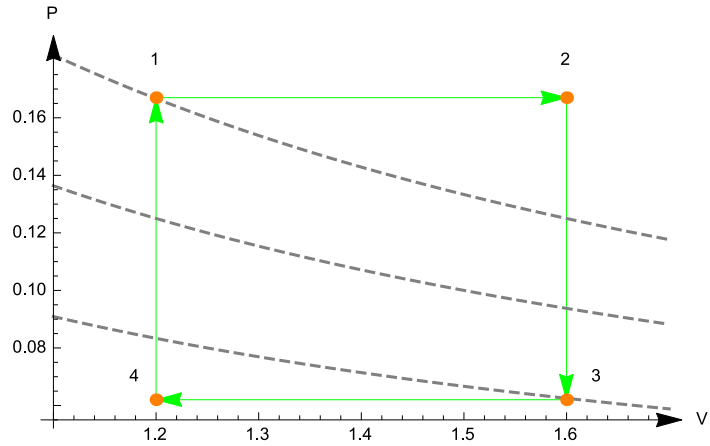}
\end{center}
 \caption{Considered cycle.} \label{fig5}
\end{figure}
\end{center}
Exploring  the equation (\ref{T}) and using the  relation  between
  the  cosmological constant and the  pressure,  we can  derive  the equation of state of the black
  holes.  The calculation gives the following equation
\begin{equation}
P=\frac{1}{4} (n-2) T \left(\frac{(n-1) V}{\omega
   }\right)^{-\frac{1}{n-1}}-\frac{\left(n^2-5 n+6\right) }{16 \pi }
\left(\frac{(n-1) V}{\omega
   }\right)^{-\frac{2}{n-1}}
   \end{equation}
where  the thermodynamical  volume $V$ is linked to the horizon
radius $r_ +$  via the relation
\begin{equation}
V=\frac{\omega}{n-1} r_+^{n-1}.
\end{equation}
Now it is possible to extract the  work of the cycle: Expressing the
volume in terms of the entropy, which will be used to reduce the
number of variables in the final expression for the efficiency, the
work takes the following form
\begin{equation}
W=\frac{4^{\frac{1}{n-2}+1}}{n-1} (P_1-P_4) \left(S_2 \left(\frac{S_2}{\omega }\right)^{\frac{1}{n-2}}-S_1
   \left(\frac{S_1}{\omega }\right)^{\frac{1}{n-2}}\right)
\end{equation}
where the subscripts refer to the quantities evaluated at the
corners labeled $(1, 2, 3, 4)$. To derive the efficiency, one has to compute the heat
quantity. The upper isobar transformation  will
produce the net inflow of  the heat which  will be identified with
$Q_H$. The latter  is expressed as follows
\begin{equation}
Q_H=\int_{T_1}^{T_2}C_p(P_1,T)dt.
\end{equation}
The integration  do not look  nice due to
 the entropy dependence of $C_p$ giving  non-trivial
$T$ dependence.  

Thus the efficiency is given by
\begin{equation}
 \eta=\frac{W}{Q_H}.
\end{equation}
To determine such a quantity, we will use    the large value limits of  $T$
and $P$. In this way,  the equation (\ref{eq13}) becomes
\begin{eqnarray}\nonumber
S&=&\frac{1}{3} 2^{\frac{-2 n^2+3 n-4}{n-2}} (n-2)^{n-5} \omega  P^{2-n} T^{n-8}
   \left[  3\ 2^{\frac{3 n}{n-2}} (n-2)^3
   T^6   -\frac{3}{\pi }\
   2^{\frac{3 n}{n-2}} (n-3) (n-2)^3 P T^4    \right.\\
   &+& \left.\frac{3 }{\pi ^2}\ 2^{\frac{n}{n-2}+1} (n-3)^2 \left(2^{\frac{4}{n-2}} n^3-9\
   2^{\frac{4}{n-2}} n^2+3\ 2^{\frac{2 n}{n-2}+1} n-5\ 2^{\frac{2 n}{n-2}}\right) P^2
   T^2\right.\\ \nonumber
   &+&\left.\frac{4}{\pi ^3} \left(-3\ 2^{\frac{2 (n+1)}{n-2}} +3\ 2^{\frac{n+4}{n-2}}
  (n-3) -2^{\frac{6}{n-2}} (n-4) (n-3) \right)(n-2) (n-3)^3 P^3+\cdots\right]
\end{eqnarray}
In this limit,  the  heat capacity (\ref{eq11})  reduces to
\begin{eqnarray}\nonumber
C_p&=&2^{\frac{-2 n^2+3 n-8}{n-2}} \omega (n-2)^{n-3}\left(4^{\frac{5}{n-2}+2} (n-5) n+3\
   32^{\frac{n}{n-2}}\right) \frac{T^{n-4}}{\pi
       P^{n-3}}\\&+&4^{1-n} (n-2) \omega  \left(\frac{(n-2)
   T}{P}\right)^{n-2}+\cdots.
\end{eqnarray}
which infers the following expression
\begin{eqnarray}\nonumber
Q_H&=&\frac{2^{1-2 n} \omega  \left(\frac{P_1}{n-2}\right)^{2-n} }{\pi ^2 (n-1) T_1^5 T_2^5}\left[(n-3)^2 (n-1) P_1^2 \left(T_1^5 T_2^n-T_2^5 T_1^n\right)+2
   \pi ^2 (n-2) T_1^4 T_2^4 \left(T_1 T_2^n-T_2 T_1^n\right)\right]+\cdot\\ \nonumber
   &=& -\frac{1}{3} 2^{\frac{-4 n^2+17 n-12}{n-2}} \left(n^2-5 n+6\right)^{n-3} \pi ^{3-n}
   \omega  P_1^{4-n} \left(\frac{S_2}{\omega }\right)^{\frac{7}{n-2}} \\ \nonumber &\times&\left( \pi  2^{\frac{4}{n-2}+4} P_1
   \left(\left(\frac{S_1}{S_2}\right)^{\frac{7}{n-2}}
   \left(\frac{S_1}{\omega }\right)^{\frac{n}{2-n}}-\left(\frac{S_2}{\omega
   }\right)^{\frac{n}{2-n}}\right)\right.\\
   &+& \left.3 (n-2) \left(\frac{S_2}{\omega
   }\right)^{-\frac{2}{n-2}}
   \left(\left(\frac{S_1}{S_2}\right)^{\frac{5}{n-2}}
   \left(\frac{S_1}{\omega }\right)^{\frac{n}{2-n}}-\left(\frac{S_2}{\omega
   }\right)^{\frac{n}{2-n}}\right)+\cdots.
 \right).
\end{eqnarray}

Then,  the efficiency is finally given by
 
\begin{eqnarray}\nonumber
\eta&=&3\ 2^{\frac{2 n}{n-2}-\frac{20}{n-2}-19} (n-3)^3 (n-2)^{3-n} \pi ^{n-5} P_1^{n-6}
   (P_1-P_4) \left(\left(\frac{S_2}{\omega
   }\right)^{\frac{n-1}{n-2}}-\left(\frac{S_1}{\omega
   }\right)^{\frac{n-1}{n-2}}\right)  \left(\frac{S_2}{\omega }\right)^{-\frac{7}{n-2}}\\ \nonumber
   &\times&\left[3 (n-2) \left(\frac{S_2}{\omega }\right)^{-\frac{2}{n-2}}
   \left(\left(\frac{S_1}{S_2}\right)^{\frac{5}{n-2}}
   \left(4^{\frac{1}{2-n}} (n-3) \left(\frac{S_1}{\omega
   }\right)^{\frac{1}{2-n}}\right)^n-\left(4^{\frac{1}{2-n}} (n-3)
   \left(\frac{S_2}{\omega }\right)^{\frac{1}{2-n}}\right)^n\right)\right]\\ \nonumber
   &-&\left.\pi
   2^{\frac{4}{n-2}+4} P_1
   \left(\left(\frac{S_1}{S_2}\right)^{\frac{7}{n-2}}
   \left(4^{\frac{1}{2-n}} (n-3) \left(\frac{S_1}{\omega
   }\right)^{\frac{1}{2-n}}\right)^n-\left(4^{\frac{1}{2-n}} (n-3)
   \left(\frac{S_2}{\omega }\right)^{\frac{1}{2-n}}\right)^n\right) \right]\\\nonumber
 &\times&\left[  (n-1) \left(\left(2^{-\frac{2}{n-2}-2} (n-3) \left(\frac{S_2}{\omega
   }\right)^{\frac{1}{2-n}}\right)^n-\left(\frac{S_1}{S_2}\right)^{\frac{7}{
   n-2}} \left(2^{-\frac{2}{n-2}-2} (n-3) \left(\frac{S_1}{\omega
   }\right)^{\frac{1}{2-n}}\right)^n\right)^2\right]^{-1}  +\cdots\\
\end{eqnarray}

The efficiency can be calculated at leading order:     Identifying  $T_C =T_4$ and $T_H =T_2$, corresponding to the lowest and
highest temperatures of the engine,  with $P\sim \frac{1}{4} (n-2) T \left(\frac{(n-1) V}{\omega
   }\right)^{-\frac{1}{n-1}}+\cdots$, we obtain
\begin{equation}
\eta=1-\frac{T_C}{T_H}\left(\frac{V_2}{V_4}\right)^{\frac{1}{n-1}}.
\end{equation}
In this way,   we can compare with the efficiency of the
Carnot  cycle $\eta_{Carnot}=1-\frac{T_C}{T_H}$.  Again, it is worth noting that
for $n=4$,  we recover the result reported in \cite{Cliff}.

\section{Conclusion}
 In this work, we have investigated the heat properties of AdS black holes in higher
dimensions.  We  have considered  the study of
corresponding thermodynamical properties along with the sign of the
heat capacity explored in the determination of the stability of
such black hole. More precisely,   we have computed the latent heat  as a trivial consequence of the Hawking-Page phase
transition.  To overcome the instability problem, the Maxwell'
construction  have been elaborated to  modify the Hawking-Page phase
structure in the $(T,S)$ plane.  We have derived  the equal
 area isotherm for any dimension in the range $4\leq n\leq 11$. Then,  we have  analyzed
  the   thermodynamic cycle and the holographic heat engines using  the expression of
 the extracted work and efficiency.

By following \cite{Cliff}, it is  possible to  make  contact  with
Maldacena conjecture known by  $AdS/CFT$ holographic correspondence
\cite{witten,cliff24}.   This could be useful   to bring new
features in the gauge theories embedded in string theory and related
topics.

\end{document}